\documentclass[aps,prl,twocolumn,superscriptaddress,showpacs]{revtex4}
\usepackage{graphicx}
\usepackage{color}

\begin{document}

\title{Electromagnetic M1 Transition Strengths from Inelastic Proton Scattering: The Cases of $^{48}$Ca and $^{208}$Pb}

\newcommand{\RCNP}{Research Center for Nuclear Physics, Osaka University, Ibaraki, Osaka 567-0047, Japan}
\newcommand{\TUDarmstadt}{Institut f\"ur Kernphysik, Technische Universit\"{a}t Darmstadt, D-64289 Darmstadt, Germany}
\newcommand{\ECT}{ECT$^{\ast}$, Villa Tambosi, I-38123 Villazzano (Trento), Italy}
\newcommand{\TWMU}{Tokyo Women's Medical University, 8-1 Kawada-cho, Shinjuku-ku, Tokyo 162-8666, Japan}

\author{J.~Birkhan}\affiliation{\TUDarmstadt}
\author{H.~Matsubara}\affiliation{\RCNP}\affiliation{\TWMU}
\author{P.~von~Neumann-Cosel}\email{vnc@ikp.tu-darmstadt.de}\affiliation{\TUDarmstadt}
\author{N.~Pietralla}\affiliation{\TUDarmstadt}
\author{V.~Yu.~Ponomarev}\affiliation{\TUDarmstadt}
\author{A.~Richter}\affiliation{\TUDarmstadt}
\author{A.~Tamii}\affiliation{\RCNP}
\author{J.~Wambach}\affiliation{\TUDarmstadt}\affiliation{\ECT}

\date{\today}

\begin{abstract}
Inelastic proton scattering at energies of a few hundred MeV and extreme forward angles selectively excites the isovector spin-flip M1 (IVSM1) resonance.
A method based on isospin symmetry is presented to extract its electromagnetic transition strength from the $(p,p')$ cross sections.
It is applied to $^{48}$Ca, a key case for an interpretation of the quenching phenomenon of the spin-isospin response, and leads to a M1 strength consistent with an older $(e,e')$ experiment excluding the almost two times larger value from a recent $(\gamma,n)$ experiment.    
Good agreement with electromagnetic probes is observed in $^{208}$Pb suggesting the possibility to extract systematic information on the IVSM1 resonance in heavy nuclei.
\end{abstract}


\maketitle

{\em Introduction.}--The isovector spin-flip M1 (IVSM1) resonance is a fundamental excitation mode of nuclei \cite{hey10}.
Its properties impact on diverse fields like the description of neutral-current neutrino interactions in supernovae \cite{lan04,lan08}, $\gamma$ strength functions utilized for physics of reactor design \cite{cha12} or for modeling of reaction cross sections in large-scale nucleosynthesis network calculations \cite{loe12}, and the evolution of single-particle properties leading to new shell closures in neutron-rich nuclei \cite{ots05}.
It also contributes to the long-standing problem of quenching of the spin-isopin response in nuclei \cite{ost92}, whose understanding is, e.g., a prerequisite for reliable calculations of nuclear matrix elements needed to determine absolute neutrino masses from a positive neutrinoless double $\beta$ decay experiment \cite{ver12}.
         
The strength distributions of the IVSM1 resonance in light and medium-mass ($fp$-shell) nuclei have been studied extensively using electromagnetic probes like electron scattering and nuclear resonance fluorescence (NRF).
However, information in heavy nuclei is limited to a few magic nuclei \cite{las86,las87,ala89,ton12,rus13}, and it is questionable whether the full strength has been observed since NRF is typically applicable only up to the neutron threshold.
Furthermore, there is no model-independent sum rule for the IVSM1 resonance like in the case of electric or Gamow-Teller (GT) giant resonances.
One exception is $^{208}$Pb, where additional information from neutron resonance studies above threshold is available \cite{koh87} and observation of the complete M1 strength distribution is claimed \cite{las88}.

The $J^\pi = 1^+$ states forming the IVSM1 resonance in even-even nuclei can also be excited in small-angle inelastic proton scattering at energies of a few hundred MeV because angular momentum transfer $\Delta L = 0$ is favored in these kinematics and the spin-isospin part dominates the proton-nucleus interaction leading to the population of the IVSM1 mode.
Indeed, in pioneering experiments bumps were observed in forward-angle scattering spectra and identified as IVSM1 resonance in heavy nuclei \cite{dja82,fre90}.    
At energies above 100 MeV  a single-step reaction mechanism dominates in $(p,p')$ scattering in analogy to the $(p,n)$ and $(n,p)$ reactions \cite{ich06}.  
It allows to relate the measured cross sections to the transition matrix elements.     
However, the classical extraction depends on model wave functions of the initial and final states and on the description of the projectile-target interaction, leading to large uncertainties. 

It is the aim of this letter to present a new method for the extraction of electromagnetic M1 transition strength from such $(p,p')$ experiments based on isospin symmetry of the IVSM1 mode and the analog GT mode excited in charge-exchange (CE) reactions.
The connection between the M1 and GT modes by isospin symmetry has been discussed extensively \cite{ost92,fuj11} and used e.g.\ to determine isospin quantum numbers of $1^+$ states from combined $(p,p')$ and $(^3{\rm He},t)$ experiments on the same target nucleus \cite{fuj07} or to derive $B$(M1) strengths from the GT matrix elements \cite{ada12}.
In CE reactions, the GT strength is obtained from normalization of the cross sections to $\beta$ decay by the so-called unit cross section \cite{tad87,zeg07}.
Here, we show that the method can be extended to the $(p,p')$ reaction opening a route to systematic studies of the IVSM1 resonance in heavy nuclei.
    
The technique is applied to two cases of particular interest, $^{48}$Ca and $^{208}$Pb.
The IVSM1 resonance in $^{48}$Ca is especially simple. 
Its strength is largely concentrated in the excitation of a single state at 10.23 MeV.
It was first observed in inelastic electron scattering  \cite{ste83} with a reduced transition strength $B(\text{M1}) = 3.9(3)$~$\mu_N^2$.
Because of its simple $[\nu 1f_{7/2}^{-1} 1f_{5/2}]$ particle-hole structure, it has been a key reference for an interpretation of the phenomenon of quenching (see, e.g., Ref.~\cite{tak88} and references therein).   
Recently, a new result from a $^{48}$Ca$(\gamma,n)$ measurement at the HI$\gamma$S facility has been reported \cite{tom11}. 
The deduced strength $B(\text{M1}) = 6.8(5)$~$\mu_N^2$  is almost two times larger.
If correct, this value would question our present understanding of quenching in microscopic models. 
For example, the consistent shell-model  quenching factors of the IVSM1 (including $^{48}$Ca) \cite{vnc98} and GT $\beta$ decay strength \cite{mar96} in $fp$-shell nuclei, successfully applied to the modeling of weak interaction processes in stars \cite{lan03}, would be challenged. 
Another important case is $^{208}$Pb \cite{bro80}, the only case where the complete $B(\text{M1})$ strength distribution is claimed to be known \cite{las88}.    

{\em Experiments}.--Recently, high energy-resolution measurements of inelastic proton scattering at extreme forward angles including $0^\circ$ have become feasible \cite{tam09,nev11}.
At RCNP Osaka, Japan, experiments have been performed at an incident proton energy of 295 MeV covering a wide range of nuclei including $^{208}$Pb.
It was demonstrated by two independent methods based on spin transfer observables and a multipole decomposition analysis (MDA) of angular distributions that the cross sections due to excitation of the IVSM1 resonance can be extracted \cite{tam11,pol12,kru15,has15}.
A corresponding $^{48}$Ca$(p,p')$ experiment was performed with a beam intensity of $4 -10$ nA. 
Protons were scattered off a $^{48}$Ca foil with an isotopic enrichment of 95.2 \% and an areal density of 1.87 mg/cm$^2$. 
Data were taken with the Grand Raiden spectrometer \cite{fuj99} in the laboratory scattering angle range $0^\circ - 5.5^\circ$ for excitation energies $5 - 25$ MeV. 
Dispersion matching techniques were applied to achieve an energy resolution of about 25 keV (full width at half maximum).
Details of the experimental techniques and the data analysis are described in Ref.~\cite{tam09}. 

\begin{figure}[tbh]
\includegraphics[width=8cm]{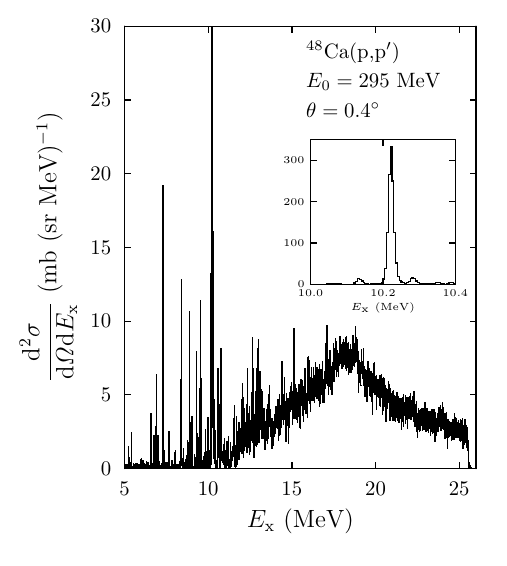}
\caption{
Spectrum of the $^{48}$Ca$(p,p')$ reaction at $E_0 = 295$~MeV and $\theta = 0.4^\circ$.
The inset shows the spectral region in the vicinity of the dominating transition at $E_{\rm x} =10.23$~MeV. 
Note the factor of ten difference in the Y-axis.
\label{Ca48-M1-fig1}}
\end{figure}

\begin{figure}[tbh]
\includegraphics[width=8cm]{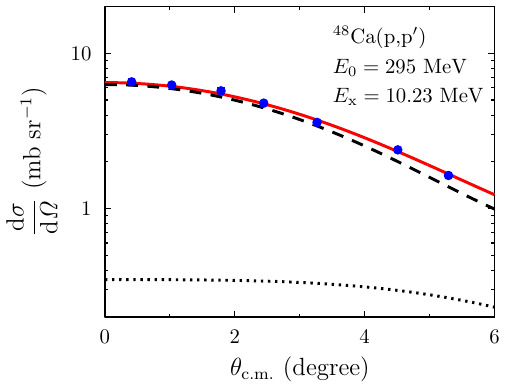}
\caption{(color online).
Angular distribution of the peak at $E_{\rm x} =10.23$~MeV (full circles) excited in the $^{48}$Ca$(p,p’)$ reaction in comparison to model calculations with the code DWBA07 for a neutron (solid line), isoscalar (dotted line), and isovector (dashed line) spin-flip $1f_{7/2}$$\rightarrow$$1f_{5/2}$ transition and the Love-Franey effective proton-nucleus interaction \cite{lov81,fra85}. 
\label{Ca48-M1-fig2}}
\end{figure}

The excitation of the $1^+$ state at 10.23 MeV is by far the strongest line in all spectra as shown by way of example for $\theta = 0.4^\circ$ in Fig.~\ref{Ca48-M1-fig1}.
In these kinematics relativistic Coulomb excitation of $J^\pi = 1^-$ states dominates the $(p,p')$ cross sections \cite{tam11,pol12,kru15,has15}.   
The broad structure peaking at about 18.5 MeV is identified as the isovector electric giant dipole resonance consistent with data from a $^{48}$Ca$(e,e'n)$ experiment \cite{str00}.
The inset of Fig.~\ref{Ca48-M1-fig1} shows an expanded spectrum around the peak at 10.23 MeV.  
Clearly, in this energy region the spectrum is free of background and a separation from other close-lying transitions is easily achieved.
Its angular distribution is presented in Fig.~\ref{Ca48-M1-fig2} (full circles). 
In order to prove the $\Delta L = 0$, spin-flip character of the transition it is compared to a theoretical angular distribution (solid line) calculated with the code DWBA07~\cite{ray07}  assuming a  neutron spin-flip $1f_{7/2}$$\rightarrow$$1f_{5/2}$ transition and using the Love-Franey effective proton-nucleus interaction~\cite{lov81,fra85}. 

{\em Extraction of M1 strength}.--While the forward-peaked $\Delta L = 0$ angular distribution can be well described independent of details of the DWBA calculation, absolute predictions of cross sections show a large uncertainty depending on the choice of the effective proton-nucleus interaction \cite{hof07}.  
Therefore, in the following we employ the concept of unit cross section developed for the extraction of GT strength from CE reactions \cite{tad87,zeg07} and derive a similar relation for the $(p,p')$ reaction.
For CE reactions the cross section at scattering angle $\theta =0^\circ$ can be written as
\begin{equation}
\frac{\mathrm{d}\sigma}{\mathrm{d}\Omega}({\rm CE}, 0^\circ) = \hat{\sigma}_\mathrm{GT} F(q,\omega) B(\mathrm{GT}), 
\label{eqgt}
\end{equation}
where $\hat{\sigma}_\mathrm{GT}$ is a nuclear-mass dependent factor (the unit cross section), $F(q,\omega)$ a kinematical factor normalized to $F(0,0)=1$ correcting for non-zero momentum and energy transfer, and $B$(GT) the reduced GT transition strength. 
The total energy transfer $\omega = E_{\rm x} - Q$, where $Q$ denotes the reaction $Q$ value.
One can define a corresponding relation for the inelastic scattering cross sections 
\begin{equation}
\frac{\mathrm{d}\sigma}{\mathrm{d}\Omega}(p,p', 0^\circ) = \hat{\sigma}_\mathrm{M1} F(q,E_x) B(\mathrm{M1}_{\sigma \tau}), 
\label{eqm1} 
\end{equation}
where $B(\mathrm{M1}_{\sigma\tau})$ denotes the reduced IVSM1 transition strength.
The kinematical correction factor is determined by DWBA calculations and the extrapolation to the cross section at $0^\circ$ from experimental data at finite angles is achieved with the aid of theoretical angular distributions as shown in Fig.~\ref{Ca48-M1-fig2}. 

The reduced GT and IVSM1 transition strengths from a $J^\pi = 0^+$ ground state to a $J^\pi = 1^+$ excited state 
can be expressed as
\begin{eqnarray}
B(\mathrm{GT})  & = & \frac{C_{\mathrm{GT}}^2}{2(2T_f+1)}  |\langle f||| \sum_k^A \sigma_k \tau_k ||| i \rangle |^2  \\
B(\mathrm{M1}_{\sigma \tau}) & = & \frac{C_{\mathrm{M1}}^2}{4(2T_f+1)}  |\langle f||| \sum_k^A \sigma_k \tau_k ||| i \rangle |^2.\label{eqme}
\end{eqnarray}
Here, $\sigma_k$ and $\tau_k$ are the spin and isospin operators acting on the $k^{\rm th}$  nucleon, $\langle|||\sigma \tau||| \rangle$ denotes a matrix element reduced in spin and isospin, and $i,f$ are initial and final states with isospin $T_i,T_f$.
The Clebsch-Gordan coefficients $C_{\mathrm{GT/M1}}$ depend on the reaction and on the $T_i$,$T_f$ values \cite{fuj11}. 
The $(p,n)$ reaction can excite GT transitions to states with isospin $T_f = T_i-1,T_i,T_i+1$ and the corresponding strength is commonly termed $B(\mathrm{GT}_-)$, $B(\mathrm{GT}_0)$, $B(\mathrm{GT}_+)$. 
The $\beta$ decay transitions used to determine the parameters of Eq.~(\ref{equnitgt}) possess $T_f = T_i -1$ while the IVSM1 resonance observed in the $(p,p')$ reaction has $T_f = T_i$.
(We note that $T_i + 1$ states can also be excited but are well separated in excitation energy and are strongly suppressed for large values of $T_i$ \cite{fuj11}).

At the very small momentum transfers considered here, isospin symmetry predicts $\hat{\sigma}_{\mathrm{M1}} \simeq \hat{\sigma}_\mathrm{GT}$.
The systematics of $\hat{\sigma}_\mathrm{GT}$ for the $(p,n)$ reaction at 297 MeV has been studied in Ref.~\cite{sas09}. 
A parameterization of its mass dependence
\begin{equation}
\hat{\sigma}_\mathrm{GT} = 3.4(3)\mathrm{exp} \left[ -0.40(5) \left( A^{1/3} - 90^{1/3} \right) \right],
\label{equnitgt} 
\end{equation} 
allows to extract $\hat{\sigma}_{\mathrm{M1}}$ for $^{48}$Ca and $^{208}$Pb.
The mass dependence of Eq.~(\ref{equnitgt}) is in very good agreement with a recent analysis of $\hat{\sigma}_{\mathrm{M1}}$ in lighter nuclei \cite{mat15}. 
The assumption of equal unit cross sections leads to
\begin{equation} 
B({\mathrm M1_{\sigma \tau}}) = \frac{1}{2} \frac{T_i}{T_i + 1}  B({\mathrm GT}_-) 
\label{eqratio1}
\end{equation}
and for the case of an analog GT transition with $T_f = T_i$ 
\begin{equation} 
B({\mathrm M1_{\sigma \tau}}) = \frac{1}{2} T_i  B({\mathrm GT}_0). 
\label{eqratio2}
\end{equation}
Equations (\ref{eqratio1},\ref{eqratio2}) imply that the IVSM1 matrix elements can also be dervied from studies of the GT strength with the $(p,n)$ reaction in the same kinematics.


The corresponding electromagnetic $B$(M1) transition strength
\begin{equation}
B(\mathrm{M1}) = \frac{3}{4\pi} |\langle f|| g_l^{\mathrm{IS}}\vec{l} + \frac{g_s^{\mathrm{IS}}}{2}\vec{\sigma} - (g_l^{\mathrm{IV}} \vec{l} + \frac{g_s^{\mathrm{IV}}}{2}\vec{\sigma})\tau_0||i \rangle |^2\, \mu_\mathrm{N}^2
\label{eqbm1em}
\end{equation}
contains spin and orbital contributions for the isoscalar (IS) and isovector (IV) parts.
For small orbital and IS contributions $B(\mathrm{M1})$ and $B({\mathrm M1_{\sigma \tau}})$ can be related by  
\begin{equation}
B(\mathrm{M1}) \cong  \frac{3}{4\pi}\left(g_s^{\mathrm{IV}} \right)^2 B(\mathrm{M1}_{\sigma \tau}) \, \mu_\mathrm{N}^2.
\label{eqbm1pp}
\end{equation}

A number of approximations is made in the derivation of Eqs.~(\ref{eqratio1},\ref{eqratio2},\ref{eqbm1pp}) whose validity is discussed in the following.
Several effects can break the equality of Eqs.~(\ref{eqratio1},\ref{eqratio2}).
In contrast to the purely IV CE reactions, the $(p,p')$ cross sections contain IS contributions.
However, because of the dominance of the $\sigma\tau$ over the $\sigma$ part of the effective interaction \cite{lov81} these are typically $\leq 5$\% and energetically separated in heavy nuclei \cite{hey10}.
Differences of exchange terms contributing to the $(p,p')$ and $(p,n)$ cross sections and Coulomb effects lead to negligible effects in the extrapolation of cross sections to $q = 0$.
A general problem of the $(p,p')$ as well as CE reactions are incoherent and coherent $\Delta L = 2$ contributions to the excitation of $1^+$ states, the latter due to the tensor part of the interaction.
Because of the difference of angular distribution shapes the incoherent $\Delta L=2$ cross sections are effectively taken into account in the MDA of the data, while the coherent part requires explicit knowledge of the excited-state wave function.
In Ref.~\cite{zeg06} a shell-model study has been performed indicating $10 - 20$\% changes of individual transition strengths with decreasing importance for stronger transitions and random sign. Thus, for the total strength the uncertainties should be smaller than 10\%.

Going from Eq.~(\ref{eqbm1em}) to Eq.~(\ref{eqbm1pp}) is justified by the following arguments:
Because of the anomalous proton and neutron $g$ factors the IS spin part is small [$(g_s^{\rm IS})^2 \approx 0.035 (g_s^{\rm IV})^2$] and can usually be neglected (see, however, the special case of $^{48}$Ca discussed below).
Furthermore, orbital M1 strength is related to deformation \cite{hey10} and disappears in the closed-shell nuclei studied in the present work.
However, Eq.~(\ref{eqbm1pp}) should approximately hold in general.
For light deformed nuclei the spin-orbital interference can be sizable for indivdual transitions but the overall strength is weakly modified ($\leq 10$\%) again because of the random mixing sign \cite{fay97,ric90}.
In heavy deformed nuclei, spin and orbital M1 strengths are energetically well separated and mixing is predicted to be weak \cite{hey10}.

Finally, meson exchange current contributions can differ for electromagnetic and hadronic reactions.
These differences are relevant in light nuclei and have e.g.\ been observed in the comparison of M1 and GT strengths in $sd$-shell nuclei \cite{ric90}.
However, for $A \geq 40$ the available data indicate that the quenching factors in microscopic calculations are the same \cite{ric85}, consistent with theoretical expectations \cite{tok80,tow87}.    

\begin{figure}[tbh]
\includegraphics[width=8cm]{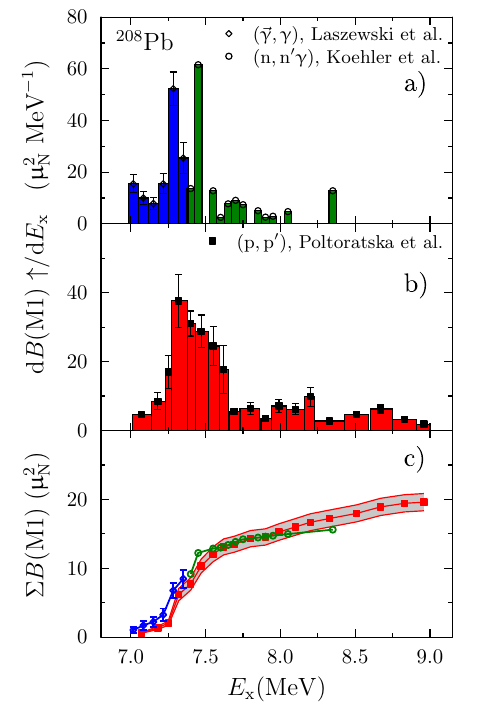}
\caption{(color online).
$B$(M1) strength distribution in $^{208}$Pb between 6.5 and 9 MeV from  (a) Refs.~\cite{las88,koh87}  and (b) from the  M1 proton scattering cross sections of Ref.~\cite{pol12} applying the method described in thepresent work. 
(c) Comparison of running sums.
\label{Ca48-M1-fig3}}
\end{figure}

{\em The case of $^{\it 208}\!$Pb.}--
We can test the approach for $^{208}$Pb, where information on the M1 strength is claimed to be complete \cite{las88}.
Figure~\ref {Ca48-M1-fig3}(a) presents the combined data of $(\vec{\gamma},\gamma')$ \cite{las88} and  $(n,n'\gamma)$ \cite{koh87} experiments providing information below and above threshold, respectively.
(Note that the strength below 7 MeV quoted in \cite{las88} is not considered because it has error bars close to 100\% and is excluded by subsequent NRF experiments \cite{rye02,shi08,sch10}).
The $B(\mathrm{M1})$ strength distribution extracted from the $(p,p')$ cross sections \cite{pol12} is presented in  Fig.~\ref {Ca48-M1-fig3}(b).
The agreement of the energy distribution and total strength is excellent [the seeming discrepancies around 7.5 MeV result from the different binning of the two data sets, cf.\ Fig.~\ref {Ca48-M1-fig3}(c)].
For example, the summed strength up to 8 MeV in Ref.~\cite{las88} of $14.8^{+1.9}_{-1.5}~\mu_\mathrm{N}^2$  is to be compared with 16.0(1.2)~$\mu_\mathrm{N}^2$ from the $(p,p')$ data. 
In the energy region between 8 and 9 MeV, previous experiments had limited sensitivity (cf.\ Fig.~6 in Ref.~\cite{koh87}), which explains why the strength seen in the $(p,p')$ experiment was missed.
In the present work, we find a total strength $\sum B(\mathrm{M1})= 20.5(1.3)$~$\mu_\mathrm{N}^2$ for the spin-M1 resonance in $^{208}$Pb.


%

{\em The case of $^{\it 48}\!$Ca}.-- The strong transition in $^{48}$Ca has pure neutron character \cite{deh84}.
In this particular case the $\vec{\sigma}$ term in the electromagnetic operator, Eq.~(\ref{eqbm1em}), cannot be neglected in the extraction of the $B(\mathrm{M1})$ value because of the interference term.   
The IS contribution to the $(p,p^\prime)$ cross sections was estimated using theoretical angular distributions for IS and IV $1f_{7/2} \rightarrow 1f_{5/2}$ transitions shown in Fig.~\ref{Ca48-M1-fig2} as dashed and dotted lines, respectively. 
A $\chi^2$ fit yields that 94.8(25)\% of the cross section at $0^\circ$ is of IV nature and the corresponding $B(\mathrm{M1}_{\sigma \tau})$ strength is deduced from Eq.~(\ref{eqm1}).

\begin{figure}[tbh]
\includegraphics[width=8cm]{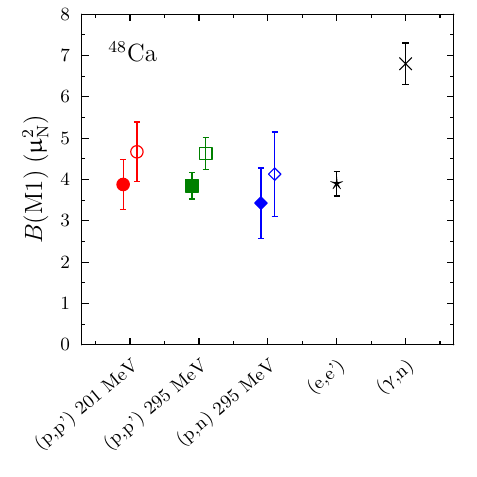}
\caption{(color online).
$B$(M1) strengths for the transition to the 10.23 MeV state in $^{48}$Ca deduced from different experiments.
The dependence on the unknown quenching of the IS part in the hadronic reactions is illustrated assuming no quenching (full symbols) or taking the value for IV quenching (open symbols).  
\label{Ca48-M1-fig4}}
\end{figure}

Extraction of the analog electromagnetic strength requires the inclusion of quenching conveniently implemented in microscopic calculations by effective  $g$ factors $g_{s,\mathrm{eff}}^{\rm IS/IV} = q^{\rm IS/IV} \times g_{s}^{\rm IS/IV}$ in Eq.~(\ref{eqbm1em}), where $q$ denotes the magnitude of quenching. 
For the IV strength a quenching factor $q^{\rm IV} = 0.75(2)$ for $fp$-shell nuclei was determined in Ref.~\cite{vnc98} and one may assume $q^{\rm IS} =q^{\rm IV}$ for the IS part. 
However, it is generally expected that ISSM1 strength is less quenched \cite{lip84}.
A recent study  in a series of $sd$-shell nuclei indicates that shell-model calculations can describe the ISSM1 strength without the need for a quenching factor \cite{mat15}, i.e.\ $g^{\mathrm{IS}}_{s,\mathrm{eff}} = g^{\mathrm{IS}}_s$.
Taking these two extremes one gets a range of possible transition strengths $B(\mathrm{M1}) = 3.85(32) - 4.63(38)$~$\mu_\mathrm{N}^2$.
We have applied the same analysis to older data for the $^{48}$Ca$(p,p^\prime)$ reaction at $E_0 = 200$ MeV \cite{cra83} with very similar results, see Fig.~\ref{Ca48-M1-fig4}.

With the aid of Eq.~(\ref{eqratio2}), the $B(\mathrm{M1})$ strength can also be derived from the $^{48}$Ca$(p,n)$ reaction.     
Yako {\it et al.}~\cite{yak09} investigated the GT strength distribution in $^{48}$Sc with the $^{48}$Ca$(p,n)$ reaction at the same incident energy of 295 MeV. 
The isobaric analog state of the level at 10.23 MeV in $^{48}$Ca is prominently excited at 16.84 MeV \cite{gre07} in the forward angle spectra (see Fig.~1 of Ref.~\cite{yak09}) and $B(\mathrm{M1})$ strengths ranging from 3.45(85)  to 4.1(1.0) $\mu_\mathrm{N}^2$ for the two extremes of IS quenching are extracted.
 
%
%

%
Figure \ref{Ca48-M1-fig4} summarizes the findings of the above analysis.
The $B(M1)$ strengths deduced from all three hadronic reactions agree well with each other and with the result from the $(e,e^\prime)$ experiment, in particular if no or little IS quenching is assumed.
Even considering the uncertainty due to the unknown magnitude of IS quenching, the large value from the $(\gamma,n)$ experiment is inconsistent with the present results.

{\em Summary}.--A new method for the extraction of $B$(M1) strength from inelastic proton scattering at forward angles is presented. 
Application to $^{208}$Pb shows good agreement with electromagnetic probes and highlights the sensitivity above the neutron threshold, where sizable additional strength is found not accessible in previous work.
The assumptions underlying the method are shown to be well justified by the case of the prominent transition in $^{48}$Ca, where a direct comparison with the analog transition excited in the $(p,n)$ reaction is possible. 
The $B(\mathrm{M1})$ strengths deduced from the $(p,p^\prime)$ and $(p,n)$ data agree with each other and with the $(e,e')$ result \cite{ste83}.
The much larger value from a recent $(\gamma,n)$ experiment \cite{tom11} is clearly in conflict with the $(e,e')$, $(p,p')$, and $(p,n)$ results derived by completely independent methods.
Systematic studies with this new experimental tool are under way (including a detailed comparison with results from electromagnetic probes for $^{90}$Zr \cite{las87,rus13,iwa12}) and promise for the first time a systematic picture of the IVSM1 resonance over wide mass and deformation ranges.   

We are indebted to the RCNP acccelerator crew for providing excellent beams.
K.~Yako kindly provided us with the data of Ref.~\cite{yak09}.
Discussions with B.~A.~Brown are gratefully acknowledged.
This work has been supported by the DFG (Grant SFB 1245) and the JSPS (Grants 07454051 and 14740154).


\begin{thebibliography}{abc99x}

\bibitem{hey10}
K. Heyde, P. von Neumann-Cosel, and A. Richter, 
%
Rev. Mod. Phys. {\bf 82}, 2365 (2010).

\bibitem{lan04}
K. Langanke, G. Martínez-Pinedo, P. von Neumann-Cosel, and A. Richter,
Phys. Rev. Lett. {\bf 93}, 202501 (2004). 

\bibitem{lan08}
K. Langanke, G. Mart\'{i}nez-Pinedo, B. M\"uller, H.-Th. Janka, A. Marek, W. R. Hix, A. Juodagalvis, and J. M. Sampaio,
%
Phys. Rev. Lett. {\bf 100}, 011101 (2008).

\bibitem{cha12}
M. B. Chadwick {\it et al.}, 
%
Nucl. Data Sheets {\bf 112}, 2887 (2011).

\bibitem{loe12}
H. P. Loens, K. Langanke, G. Mart\'{i}nez-Pinedo, and K. Sieja,
Eur. Phys. J. A {\bf 48}, 34 (2012).

\bibitem{ots05}
T. Otsuka, T. Suzuki, R. Fujimoto, H. Grawe, and Y. Akaishi,
%
Phys. Rev. Lett. {\bf 95}, 232502 (2005); 
%
T. Otsuka, T. Suzuki, M. Honma, Y. Utsuno, N. Tsunoda, K. Tsukiyama, and M. Hjorth-Jensen,
{\it ibid} {\bf 104}, 012501 (2010). 

\bibitem{ost92}
F. Osterfeld,
%
Rev. Mod. Phys. {\bf 64}, 491 (1992).

\bibitem{ver12}
J. D. Vergados, H. Ejiri, and F. Simkovic,
%
Rep. Prog. Phys. {\bf 75}, 103601 (2012).

\bibitem{las86}
R. M. Laszewski, P. Rullhusen, S. D. Hoblit, and S. F. LeBrun,
%
Phys. Rev. C {\bf 34}, 2013(R) (1986).

\bibitem{las87}
R. M. Laszewski, R. Alarcon, and S. D. Hoblit, 
%
Phys. Rev. Lett. {\bf 59}, 431 (1987).

\bibitem{ala89}
R. Alarcon, R. M. Laszewski, and D. S. Dale, 
%
Phys. Rev. C {\bf 40}, 1097(R) (1989).

\bibitem{ton12}
A. P. Tonchev, S. L. Hammond, J. H. Kelley, E. Kwan, H. Lenske, G. Rusev, W. Tornow, and N. Tsoneva,
%
Phys. Rev. Lett. {\bf 104}, 072501 (2010).

\bibitem{rus13}
G. Rusev {\it et al.},
%
Phys. Rev. Lett. {\bf 110}, 022503 (2013).

\bibitem{koh87}
R. K\"ohler, J. A. Wartena, H. Weigmann, L. Mewissen, F. Poortmans, J. P. Theobald, and S. Raman,
%
Phys. Rev. C {\bf 35}, 1646 (1987).

\bibitem{las88}
R. M. Laszewski, R. Alarcon, D. S. Dale, and S. D. Hoblit, 
Phys. Rev. Lett. {\bf 61}, 1710 (1988).

\bibitem{dja82}
C. Djalali, N. Marty, M. Morlet, A. Willis, J. C. Jourdain, N. Anantaraman, G. M. Crawley, A. Galonsky, and P. Kitching,
Nucl. Phys.  {\bf A388}, 1 (1982).

\bibitem{fre90}
D. Frekers {\it et al.},
Phys. Lett.  B {\bf 244}, 178 (1990).

\bibitem{ich06}
M. Ichimura, H. Sakai, and T. Wakasa,
%
Prog. Part. Nucl. Phys. {\bf 56}, 446 (2006).

\bibitem{fuj11}
Y. Fujita, B. Rubio, and W. Gelletly,
%
Prog. Part. Nucl. Phys. {\bf 66}, 549 (2011).

\bibitem{fuj07}
H. Fujita {\it et al.},
%
Phys. Rev. C {\bf 75}, 034310 (2007).

\bibitem{ada12}
T. Adachi {\it et al.},
%
Phys. Rev. C {\bf 85}, 024308 (2012).

\bibitem{tad87}
T. N. Taddeucci {\it et al.}, 
%
Nucl. Phys. {\bf A469}, 125 (1987).

\bibitem{zeg07}
R. G. T. Zegers {\it et al.},
%
Phys. Rev. Lett. {\bf 99}, 202501 (2007).

\bibitem{ste83}
W. Steffen {\it et al.},
%
Phys. Lett. B {\bf 95}, 23 (1980); Nucl. Phys. {\bf A404}, 413 (1983).

\bibitem{tak88}
K. Takayanagi, K. Shimizu, and A. Arima,
%
Nucl. Phys. {\bf A481}, 313 (1988).

\bibitem{tom11}
J. R. Tompkins, C. W. Arnold, H. J. Karwowski, G. C. Rich, L. G. Sobotka, and C. R. Howell, 
Phys. Rev. C {\bf 84}, 044331 (2011).

\bibitem{vnc98}
P. von Neumann-Cosel, A. Poves, J. Retamosa, and A. Richter, 
%
Phys. Lett. B {\bf 443}, 1 (1998).

\bibitem{mar96}
G. Mart\'{i}nez-Pinedo, A. Poves, E. Caurier,  and A. P. Zuker, 
%
Phys. Rev. C {\bf 53}, 2602(R) (1996).

\bibitem{lan03}
K. Langanke and G. Mart\'{i}nez-Pinedo,
%
Rev. Mod. Phys. {\bf 75}, 819 (2003). 


%

\bibitem{bro80}
G. E. Brown and S. Raman, 
%
Comments Nucl. Part. Phys. {\bf 9}, 79 (1980);
%
R. M. Laszewski and J. Wambach,
{\it ibid} {\bf 14}, 321 (1985).

\bibitem{tam09}
A. Tamii {\it et al.}, 
%
Nucl. Instrum. Methods  A {\bf 605}, 3 (2009).

\bibitem{nev11}
R. Neveling {\it et al.},
%
Nucl. Instrum. Methods  A {\bf 654}, 29 (2011).

\bibitem{tam11}
A. Tamii {\it et al.}, 
%
Phys. Rev. Lett. {\bf 107}, 062502 (2011).

\bibitem{pol12}
I. Poltoratska {\it et al.},
%
Phys. Rev. C {\bf 85}, 041304(R) (2012).

\bibitem{kru15}
A. M. Krumbholz {\it et al.},
%
Phys. Lett. B {\bf 744}, 7 (2015).

\bibitem{has15}
T. Hashimoto {\it et al.},
%
Phys. Rev. C {\bf 92}, 031305(R) (2015).

\bibitem{fuj99}
M. Fujiwara {\it et al.}, 
%
Nucl. Instrum. Methods Phys. Res., Sect. A {\bf 422}, 488 (1999).

\bibitem{str00}
S. Strauch, P. von Neumann-Cosel, C. Rangacharyulu, A. Richter, G. Schrieder, K. Schweda, and J. Wambach,
%
Phys. Rev. Lett. {\bf 85}, 2913 (2000).

\bibitem{ray07}
J. Raynal, program DWBA07, NEA Data Service NEA1209/08.

\bibitem{lov81}
W. G. Love and M. A. Franey, 
%
Phys. Rev. C {\bf 24}, 1073 (1981).

\bibitem{fra85}
M. A. Franey and W. G. Love, 
%
Phys. Rev. C {\bf 31}, 488 (1985).

\bibitem{hof07}
F. Hofmann {\it et al.}, 
%
Phys. Rev. C {\bf 76}, 014314 (2007).

\bibitem{sas09}
M. Sasano {\it et al.},
%
Phys. Rev. C {\bf 79}, 024602 (2009).

\bibitem{mat15}
H. Matsubara {\it et al.},
%
Phys. Rev. Lett. {\bf 115}, 102501 (2015).

\bibitem{zeg06}
R. G. T. Zegers {\it et al.},
%
Phys. Rev. C  {\bf 74}, 024309 (2006).

\bibitem{fay97}
M. S. Fayache, P. von Neumann-Cosel, A. Richter, Y. Y. Sharon, and L. Zamick,
%
Nucl. Phys. A {\bf 627}, 14 (1997).

\bibitem{ric90}
A. Richter, A. Weiss, B. A. Brown, and O. H\"{a}usser,
%
Phys. Rev. Lett. {\bf 65}, 2515 (1990).

\bibitem{ric85}
A. Richter,
%
Prog. Part. Nucl. Phys. {\bf 13}, 1 (1985).

\bibitem{tok80}
H. Toki and W. Weise, 
%
Phys. Lett. B {\bf 97}, 12 (1980).

\bibitem{tow87}
I. Towner,
%
Phys. Rep. {\bf 155}, 263 (1987).

\bibitem{rye02}
N. Ryezayeva, T. Hartmann, Y. Kalmykov, H. Lenske, P. von Neumann-Cosel, V. Yu. Ponomarev, A. Richter,
A. Shevchenko, S. Volz, and J.Wambach,
%
Phys. Rev. Lett. {\bf 89}, 272502 (2002).

\bibitem{shi08}
T. Shizuma, T. Hayakawa, H. Ohgaki, H. Toyokawa, T. Komatsubara, N. Kikuzawa, A.~Tamii, and H. Nakada,
%
Phys. Rev. C {\bf 78}, 061303(R) (2008).

\bibitem{sch10}
R. Schwengner {\em et al.},
%
Phys. Rev. C {\bf 81}, 054315 (2010).

\bibitem{deh84}
D. Dehnhard {\it et al.},
%
Phys. Rev. C {\bf 30}, 242 (1984).

%

\bibitem{lip84}
E. Lipparini and A. Richter,
%
Phys. Lett. B{\bf 144},13 (1984).

\bibitem{cra83}
G.M. Crawley, N. Anantaraman, A. Galonsky, C. Djalali, N. Marty, M. Morlet, A. Willis, and J.-C. Jourdain,
%
Phys. Lett. B {\bf 127}, 322 (1983).

\bibitem{yak09}
K. Yako {\it et al.},
%
Phys. Rev. Lett. {\bf 103}, 012503 (2009).

\bibitem{gre07}
E.-W. Grewe {\it et al.},
%
Phys. Rev. C {\bf 76}, 054307 (2007);
%
D. Frekers, private communication.

%
%
%
%
%

%

%

%
%

\bibitem{iwa12}
C. Iwamoto {\it et al.},
%
Phys. Rev. Lett. {\bf 108}, 262501 (2012).

%

\end{thebibliography}
\end{document}